\UseRawInputEncoding
\documentclass[journal]{IEEEtran}
\usepackage{amsmath, bm, amsfonts,mathrsfs}
\usepackage{algorithmic,algorithm}
\usepackage{graphics,graphicx,subfigure}
\usepackage{tabularx}
\ifCLASSINFOpdf

\else

\fi

\begin{document}
\title{Scalable Framework for Deep Learning based CSI Feedback}

\author{
Liqiang Jin, Qiuping Huang, Qiubin Gao, Yongqiang Fei, Shaohui Sun
\thanks{All authors are  with China Academy of Telecommunications Technology (CATT),
Beijing, People¡¯s Republic of China,
email: {jinliqiang}@cictmobile.com}
}

\maketitle

\begin{abstract}
Deep learning (DL) based channel state information (CSI) feedback in multiple-input multiple-output (MIMO) systems recently has attracted  lots of attention from both academia and industrial.
From a practical point of views, it is huge burden to train, transfer and deploy a DL model for each parameter configuration of the base station (BS).
In this paper, we propose a scalable and flexible framework for DL based CSI feedback referred as scalable CsiNet (SCsiNet) to adapt a family of configured parameters such as feedback payloads, MIMO channel ranks, antenna numbers.
To reduce model size and training complexity, the core block with pre-processing and post-processing in SCsiNet is reused among different parameter configurations as much as possible which is totally different from configuration-orienting design.
The pre-processing and post-processing are trainable neural network layers introduced for matching input/output dimensions and probability distributions. The proposed SCsiNet is evaluated by metrics of squared generalized cosine similarity (SGCS) and user throughput (UPT) in system level simulations. Compared to existing schemes (configuration-orienting DL schemes and 3GPP Rel-16 Type-II codebook based schemes), the proposed scheme can significantly reduce mode size and achieve $2\%\sim10\%$ UPT improvement for all parameter configurations.

\end{abstract}

\begin{IEEEkeywords}
MIMO, CSI feedback, deep learning, scalable framework.
\end{IEEEkeywords}

\IEEEpeerreviewmaketitle
\section{Introduction}
Massive multiple-input multiple-output (MIMO) is a promising technology to improve spectrum efficiency and system throughput in wirelss communication systems. However,  accurate acquisition and feedback of channel state information (CSI) is the key to ensure a good performance of massive MIMO systems. In the fifth generation new radio (5G NR) system, eigenvectors (pre-coding vectors) of downlink MIMO channels are acquired based on CSI reference signals (CSI-RS) at user equipments (UEs) and return to the base station (BS) via uplink channels. To reduce feedback overheads, the correlation of eigenvectors in the frequency domain is exploited by enhanced Type II (eType II) codebooks defined in \cite{eTypeII}.

Recently deep learning (DL) based CSI feedback has attracted lots of attention from both academia and industrial due to its potential to improve feedback accuracy, reduce feedback overhead and delay  \cite{3GPP_AI}. An auto-encoder (encoder-decoder network) called CsiNet was firstly introduced by \cite{Wen0} to compress MIMO full channels where encoder part and decoder part are mainly constructed by convolutional neural networks (CNNs) with a residual structure. Simulation results show that CsiNet outperforms traditional schemes based on compressed sensing algorithms. Inspired by this pioneering work, the temporal correlation and reciprocity between the uplink and the downlink are utilized by CsiNet-LSTM \cite{TWang} and DualNet-MAG/ABS \cite{ZLiu} to further improve feedback performances. Different from above-mentioned schemes concentrating on the feedback of MIMO full channels, the EVCsiNet proposed in \cite{WLiu} considers the feedback of channel eigenvectors and shows the superiority over the scheme based on eType II codebooks.

In existing DL based schemes for CSI feedback, all DL models are designed orienting to a specific scenario and configuration. This means that the number of DL models is at least proportional to the
number of configured parameters which will significantly increase the burden of model training, model transferring and model deployment. For example, depending on the capacity of uplink channel, the network may prefer different CSI feedback payload size. The payload size ranges from tens of bits to hundreds of bits. In addition, the BS can be equipped with different number of antenna ports, e.g., 4, 8, 12, 32 etc. The network can also have preference on the number of data layers depending on channel condition. Therefore, we may need $384=16\times4\times6$ DL models for feedback overheads of $\{20,40,60,...,320\}$-bits, MIMO ranks of $\{1,2,3,4\}$, and antenna port numbers of \{4,8,12,16,24,32\} even if other wireless communication scenarios/configurations are ignored (e.g., urban macro, urban micro, rural macro, indoor hotspot). It is unaffordable for UEs to support so many DL models from storage perspective. This motivates us to design a scalable and flexible framework for DL based CSI feedback.

In this paper, we will give a scalable and flexible framework for DL based eigenvectors feedback to adapt a family of parameter configurations called scalable CsiNet (SCsiNet). Our SCsiNet is designed as a multi-branch structure and core blocks are shared/reused among all configurations to reduce model size which is totally different from conventional configuration-orienting design. We also provide training scheme for SCsiNet and a payload allocation scheme for different MIMO layers and ranks. Simulation results show that the proposed SCsiNet can achieve a better performance under the same feedback payload compared to the scheme based on eType II codebooks.

The main contribution of this paper  is summarized as follows.
\begin{itemize}
\item A scalable and flexible framework for DL based eigenvectors feedback is designed based on a multi-branch structure to adapt a family of parameter configurations, e.g., feedback payload, MIMO ranks, antenna ports.  The core block in SCsiNet is reused among all parameter configurations and all DL models can be unified into single one.
\item With this scalable and flexible framework, rank-adaptive payload allocation is introduced to ensure performances and payload requirements for different ranks.
\item To evaluate the proposed SCsiNet, we give squared generalized cosine similarity (SGCS) performances and downlink user throughput (UPT) performances in system level simulations.
\end{itemize}

\section{System model and existing scheme}
In this paper, we consider a downlink MIMO-OFDM system with $N_t$ transmit antennas at the BS and $N_r$ receive antennas at UEs. Let $H(f)\in {\mathbb C}^{N_r\times N_t}, f=1,2,...,N_{c}$ denotes channel matrix,  subband level channel eigenvectors $w(s)$ can be  expressed as
\begin{equation}
\begin{split}
&w(s)=Eig\left(R_{sb}(s)\right)\in {\mathbb C}^{N_t\times N_{ri}}\\
&R_{sb}(s) = \frac{1}{N_c/N_{sb}}\sum\limits_{f=(s-1)\cdot N_c/N_{sb}+1}^{s\cdot N_c/N_{sb}} H^H(f)H(f)
\end{split}
\end{equation}
for $s=1,2,...,N_{sb}$, where $N_c$ and $ N_{sb}$ are the numbers of sub-carriers and sub-bands respectively, and each sub-band consists of $N_c/N_{sb}$ sub-carriers. $Eig(R(s))$ denotes $N_{ri}$ eigenvectors of $R(s)$ associated with the largest $N_{ri}$ eigenvalues, $R_{sb}(s)$ is the correlation matrix of $s$-th sub-band. Similar to \cite{WLiu}, we also use SGCS to evaluate recovery accuracy of the eigenvectors, which is defined as
\begin{equation}
\rho(W_i,\hat{W}_i) = \frac{1}{N_{sb}}\sum\limits_{s=1,2,...,N_{sb}}\frac{||w^H(s,i){\hat w}(s,i)||^2}{||w(s,i)||^2||{\hat w}(s,i)||^2}
\end{equation}
where $w(s, i)$ denote the $i$-th column of $w(s)$ i.e. the eigenvector of $s$-th sub-band and $i$-th layer,
${\hat w}(s, i)$ is the reconstructed eigenvector corresponding to $w(s, i)$, $W_i=[w(1,i),w(2,i),...,w(N_{sb},i)]\in{\mathbb C}^{N_t\times N_{sb}}$.

\subsection{3GPP eType II codebook based eigenvector feedback}
For an eType II codebook based feedback, sub-band level eigenvectors $w(s)$ are first compressed to reduce their dimensions, then complex values in low dimension are quantized to a binary stream at UE. At the BS, the binary stream is dequantized  and then used to reconstruct eigenvectors.
For the $i$-th layer,  the reconstructed eigenvectors of all sub-bands can be expressed by
\begin{equation}
W_i = W_s{\tilde W}_iW_{f,i}^H\in{\mathbb C}^{N_t\times N_{sb}}
\end{equation}
where $W_s\in{\mathbb C}^{N_t\times 2L}$ is spatial beam matrix reflecting long-term and wide-band characteristics, $W_{f,i}\in{\mathbb C}^{N_{sb}\times M}$ is frequency compression matrix, and ${\tilde W}_i\in{\mathbb C}^{2L\times M}$ is the dequantized result of a binary stream. The value of $M=\lceil p\cdot N_{sb}\rceil$ is determined by frequency compression rate $p$ configured by the BS.
The spatial beam matrix $W_s$ and frequency compression matrix $W_{f,i}$ are all made up of DFT vectors, $W_s$ is shared by all layers but $W_{f,i}$ is layer-specific.

\subsection{DL based eigenvector feedback}
In \cite{WLiu}, EVCsiNet is proposed for for eigenvector feedback. This is a typical configuration-orienting design.  Let $W_1=[w(1,1),w(2,1),...,w(N_{sb},1)]\in{\mathbb C}^{N_t\times N_{sb}}$ denote eigenvectors  associated with the largest eigenvalues for all sub-bands. EVCsiNet  tries to compress and quantize $W_1$ into binary bits and reconstruct it as follows
\begin{equation}
\begin{split}
z = Q(f_{\Theta_{E}}(W)),
\hat{W} = f_{\Theta_D}(D^{-1}(z))
\end{split}
\end{equation}
where $f_{\Theta_{E}}$ and $f_{\Theta_{D}}$ denote the encoder part and the decoder part, $Q$ and $D^{-1}$ denote the quantizer and the dequantizer.
The parameters of $\Theta_E$ and $\Theta_D$ are obtained by optimizing
\begin{equation}
\{\Theta_E,\Theta_D\}=\arg\max\limits_{\Theta_E,\Theta_D}\rho(W,\hat{W})
\end{equation}
at the training phase. Considering the implementation complexity of encoder at the UE side, lightweight architecture is adopted with two fully-connected layers in the encoder part of EVCsiNet.

\section{Proposed SCsiNet}
Various payloads, ranks and antenna numbers should be supported as that for eType II codebooks in the 5G NR system.
To this end, a trivial way is configuration-orienting design but the complexity of model training, storage and transferring would be increased linearly with the number of configurations.
A alternative way is to propose a scalable and flexible framework and all DL models are unified into single one.
In this section, this framework, i.e. SCsiNet will be introduced in detail.

\subsection{SCsiNet}

\begin{figure*}[htbp]
\centering
\includegraphics[width=0.8\textwidth]{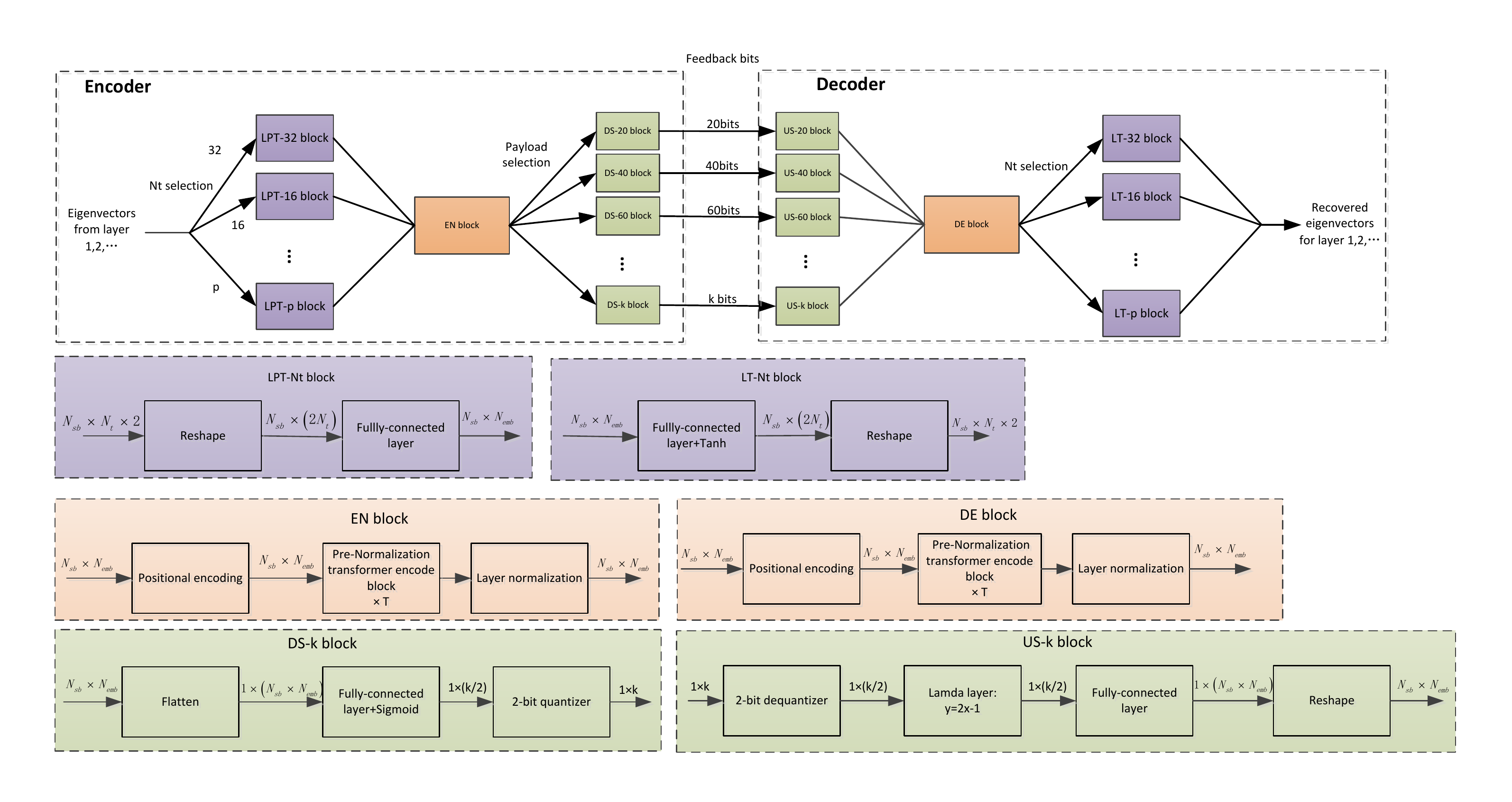}
\caption{The structure of SCsiNet.}
\label{fig:SCsiNet}
\end{figure*}

As shown in Fig.\ref{fig:SCsiNet}, the SCsiNet is a ``layer-common" model which ignores the layer index of eigenvector inputs.
In other words, for any $i=1,2,..,N_{ri}$, eigenvectors $W_i$ from $i$-th layer and $N_{sb}$ sub-bands are fed into the same SCsiNet.
To adapt a family of parameter configurations, SCsiNet is designed as a multiple branch structure where each split branch corresponds to the specific value of a parameter.
The core block (EN/DE block) has multiple branches inside or outside, and is also connected with multiple blocks (US/DS block or LPT/LT block).
These connected blocks are regarded as pre-processing or post-processing with the respect to core blocks.

Specifically, down-sampling/up-sampling (DS/US) blocks are introduced to support various
feedback payloads. DS-$k$ block and US-$k$ block are linear transformations for down-sampling and up-sampling, respectively. The output range of DS-$k$ block is restricted to $[0,1]$ with a sigmoid activation function for the convenience of quantization.
Also, 2-bits scalar uniform quantizer and dequantizer are embedded into DS-$k$ block and US-$k$ block respectively.
A pair of DS-$k$ block and US-$k$ block corresponds to a payload configuration of $k$-bits, and multiple pairs of DS-$k$ block and US-$k$ block ensure that the proposed SCsiNet is payload-scalable.
Moreover, the linear pre-transform/transform (LPT/LT) blocks are introduced for the propose of unifying input/output dimensions and probability distributions of eigenvectors from different antenna number $N_{t}\in{\mathbb N_t}$. LPT-$p$ block is a linear transformation for eigenvectors of antenna number $p$ at the spatial domain which transforms its input into a common domain while LT-$p$ block transforms its inputs back into the original eigenvector domain of antenna number $p$.
It is worth noting that all above linear transformations are realized by trainable fully-connected layers.
Two core blocks of EN block and DE block are shared among all branches, also parameter configurations.
They are made up of several transformer encode blocks with $N_{head}$ parallel attention layers, and the pre-normalization version in \cite{Nguyen} is adopted to facilitate training.

The detailed design of all blocks in SCsiNet is shown at the bottom of Fig.\ref{fig:SCsiNet}. For LPT-$p$ block, the input is eigenvectors from all sub-bands $W_i\in{\mathbb R}^{N_{sb}\times N_{t}\times 2}$ (2 means real part and imaginary part of a complex number) for  $i$-th layer, $W_i$ is linearly embedded into a higher dimension of ${\mathbb R}^{N_{sb}\times N_{emb}}$ by a fully-connected layer with $N_{emb}$ units. In LT-$p$ block, a fully-connected layer is also used with $N_{t}$ units and a tanh activation function  because the range of eigenvectors is between -1 and 1.
As mentioned above, $T_{EN}$ transformer encode blocks and $T_{DE}$ transformer encode blocks with pre-normalization are respectively used in EN block and DE block after positional encoding. The attention mechanism in transformer encode blocks can help to extract the correlation of eigenvectors among sub-bands. Since the size is not changed, the input dimension and the output dimension of EN block and DE block are  ${\mathbb R}^{N_{sb}\times N_{emb}}$. The detail of transformer encode block and positional encoding can refer to \cite{Vaswani}. In DS-$k$ block, the input of size $N_{sb}\times N_{emb}$ is firstly flatten into the dimension of $1\times (N_{emb}N_{sb})$, then reduced to the dimension of $1\times (k/2)$ by a fully-connected layer with $k/2$ units, and feedback bits are obtained by quantizing the output of this fully-connected layer. As for US-$k$ block, feedback bits are dequantized to real numbers whose range is $[0,1]$.  The ``lambda layer" is used to map the range from $[0,1]$ to $[-1,1]$ which can be described by the element-wise operation of $y=2x-1$.
Similarly, a full-connected layer with $N_{sb}\times N_{emb}$ units is used for up-sampling. The
hyper-parameters of SCsiNet are given in Table.\ref{table:hyperparameters}. For simplicity, we only consider the case of $T_{EN}=T_{DE}$ in this paper. Since the BS is more powerful,
the SCsiNet may be deigned as $T_{DE}>T_{EN}$.

\begin{table}[htbp]
\centering
\caption{Hyper-parameters of SCsiNet.}
\begin{tabular}{|c|c|c|c|c|c|c|}
\hline
\textbf{Parameter}&$N_{head}$&$N_{sb}$&$N_{ri}$&$N_{emb}$&$T_{EN}$\\
\hline
\textbf{Value}&8&12&4&128&2\\
\hline
\textbf{Parameter}&$T_{DE}$&${\mathbb K}(bits)$&$\mathbb N_t$&&\\
\hline
\textbf{Value}&2&\{20,40,60,...,320\}&\{16,32\}&&\\
\hline
\end{tabular}\label{table:hyperparameters}
\end{table}

\subsection{Training of SCsiNet}
The training process can be generally described by optimizing the following formula
\begin{equation}\label{eq:training_SCsiNet}
\{\Theta_E,\Theta_D\}=\arg\max\limits_{\Theta_E,\Theta_D}\frac{1}{16}\sum_{k=20,40,60,...,320}\rho(W_i,{\hat W}_i(k))
\end{equation}
where  ${\hat W}_i(k)$ denotes the reconstructed result of $k$-bits payload. It can be seen that given layer index $i$,
the loss function is the average SGCS over all payloads. At the training phase, the eigenvectors from all layers will be fed into SCiNet ignoring layer index $i$.
For various antenna number ${\mathbb N_t}$, a nature way to train the SCsiNet is to feed training data with different antenna number $N_t\in {\mathbb N_t}$
in turn which can be described by Algorithm.\ref{algorithm:training}.  To reduce training complexity, a three-stage training approach is used in this paper.
At the first stage, the SCsiNet is only optimized by eigenvectors from $N_t=32$ antenna number.
Then a pair of LPT-$p$ and LT-$p$ ($p\neq32$) is optimized at the second stage after EN block, DE block,
DS-$k$ block s and US-$k$ blocks are frozen (i.e., the weights are not updated by the optimizer).
Fine-tuning is adopted with few epochs based on Algorithm.\ref{algorithm:training} at the final stage.

\begin{algorithm}[htbp]
\caption{Training process of SCsiNet}
\hspace*{0.02IN}{\bf Input:}
Supported antenna numbers ${\mathbb N_t}=\{32,16...\}$,\\
training data ${\mathbb D}_{N_t}$ for $N_t\in{\mathbb N_t}$,\\
Batch size $m$\\
\begin{algorithmic}[1]
\WHILE{$\Theta_E,\Theta_D$ have not converged}
\FOR{$N_t\in {\mathbb N_t}$}
\STATE Sample $\{d_i\}_{i=1}^{m}$ a batch of data from ${\mathbb D}_{N_t}$.
\STATE Optimize SCsiNet by (\ref{eq:training_SCsiNet}) with $\{d_i\}_{i=1}^{m}$.
\ENDFOR
\ENDWHILE
\end{algorithmic}\label{algorithm:training}
\end{algorithm}

\subsection{Inference of SCsiNet}
 At the inference phase, only one branch in Fig.\ref{fig:SCsiNet} will be activated according to the configured parameter. For example, a path of ``LPT-16 block$\rightarrow$EN block$\rightarrow$DS-120 block$\rightarrow$US-120 block$\rightarrow$DE block$\rightarrow$LT-16 block" will be activated for the configuration of transmit antenna number $N_{t}=16$, 120-bits feedback payload.
For any $i=1,2,...,N_{ri}$ layer, eigenvectors of all sub-bands $W_i\in{\mathbb C}^{16\times N_{sb}}$ are fed into the encoder part of SCsiNet for compression.
Since two core blocks (EN block and DE block) are shared among all configurations and transformer encode block is usually much bigger than other blocks, the model size of the proposed SCsiNet will be superior to the configuration-orienting DL based scheme (i.e., each configuration is equipped with a DL model).

\section{Simulation results}
In this section, we give simulation results of SCsiNet, conventional configuration-orienting DL schemes and eType II codebook based schemes in system level simulations.
We adopt the channel model defined by 3GPP TR38.901 with urban macro (UMa) scenario, 2GHz carrier frequency and 15kHz subcarrier spacing.
We consider 48 physical resource blocks (PRBs) and $N_{sb}=12$ sub-bands (i.e., 4PRBs per sub-band for eigenvector feedback). The more detail of channel modeling parameters is given in Appendix.
The data set is builded by eigenvectors of antenna number $N_t=16,32$ and $N_{ri}=4$ MIMO layers from 50 drops (random seeds), 57 cells and 570 UEs (10 UEs per cell).
To improve the diversity of data set, we collect eigenvectors under the service of full buffer because
all UEs in a cell will perform channel measurement and eigenvector feedback.
The collected data set is divided into training data set and test data set according to their drops which have the size of $456000=40\times570\times20$ and $114000=10\times570\times20$ ($Drops\times UEs\times Samples$) per MIMO layer, respectively.
In fact, there are two kinds of eigenvectors in simulations as follows
\begin{itemize}
\item Ideal eigenvectors which are obtained by singular value decomposition (SVD) of known channel matrices;
\item Realistic eigenvectors which are obtained by SVD of estimated channel matrices corrupted by the noise and the interference.
\end{itemize}
In this paper, ideal eigenvectors are used for both inputs and labels at the training stage. However, at the inference stage, realistic eigenvectors are fed into DL models and the performance is evaluated by the SGCS between reconstructed eigenvectors and ideal eigenvectors. Before fed to DL models, the eigenvectors $W_i=\in{\mathbb C}^{N_t\times N_{sb}}$ will be normalized by the maximum amplitude in $W_i$, i.e. ${\bar W}_i = W_i/\max\limits_{r,c}(|W_i(r,c)|)$. At the first two training stages,  the adaptive momentum (Adam) optimizer with the learning rate of \cite{Vaswani} is adopted for 200 and 100 epochs, respectively.
 And at the fine-tuning stage, the Adam optimizer with cosine annealing learning rate of $10^{-5}\sim10^{-4}$ is adopted for 30 epochs.

\subsection{SCsiNet versus configuration-orienting DL scheme}
\begin{figure}[htbp]
\centering
\includegraphics[width=8cm]{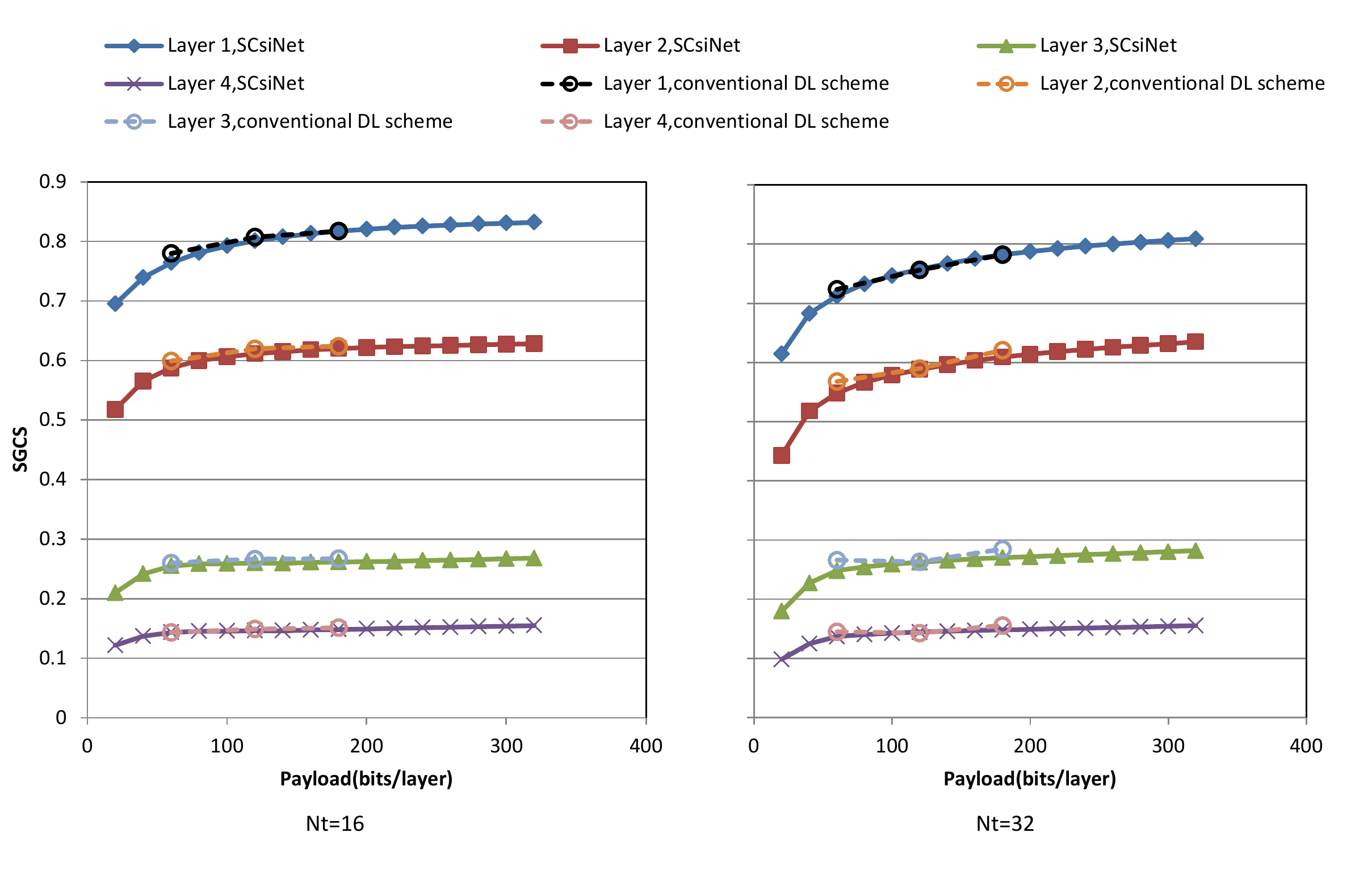}
\caption{SGCS performances of SCsiNet and conventional configuration-orienting DL scheme.}\label{fig:SGCS_SCsiNet_vs_ConventinalDL}
\end{figure}

Fig.\ref{fig:SGCS_SCsiNet_vs_ConventinalDL} shows SGCS performances of SCsiNet and conventional configuration-orienting DL schemes for various payloads and 4 layers.
In conventional DL schemes, each DL model is ``layer-common" and  trained by mixed eigenvectors from $4$ layers for a specific payload and antenna number.
In this figure, $\{60,120,180\}$-bits payloads and $\{16,32\}$ antenna numbers are considered and the total number of DL models is $6=2\times3$ for conventional DL schemes.
It can be observed from simulation results that the proposed SCsiNet has a similar performance compared to conventional DL schemes for 4 layers and different payloads.
If $\{20,40,60,...,320\}$-bits payloads and $\{16,32\}$ antenna numbers are considered, the number of DL models will even become $32=16\times2$ for conventional schemes. However, the proposed SCsiNet can adapt these payloads and antenna numbers with single DL model and small model size.

\subsection{SCsiNet versus eType II codebook based scheme}
\begin{figure}[htbp]
\centering
\includegraphics[width=8cm]{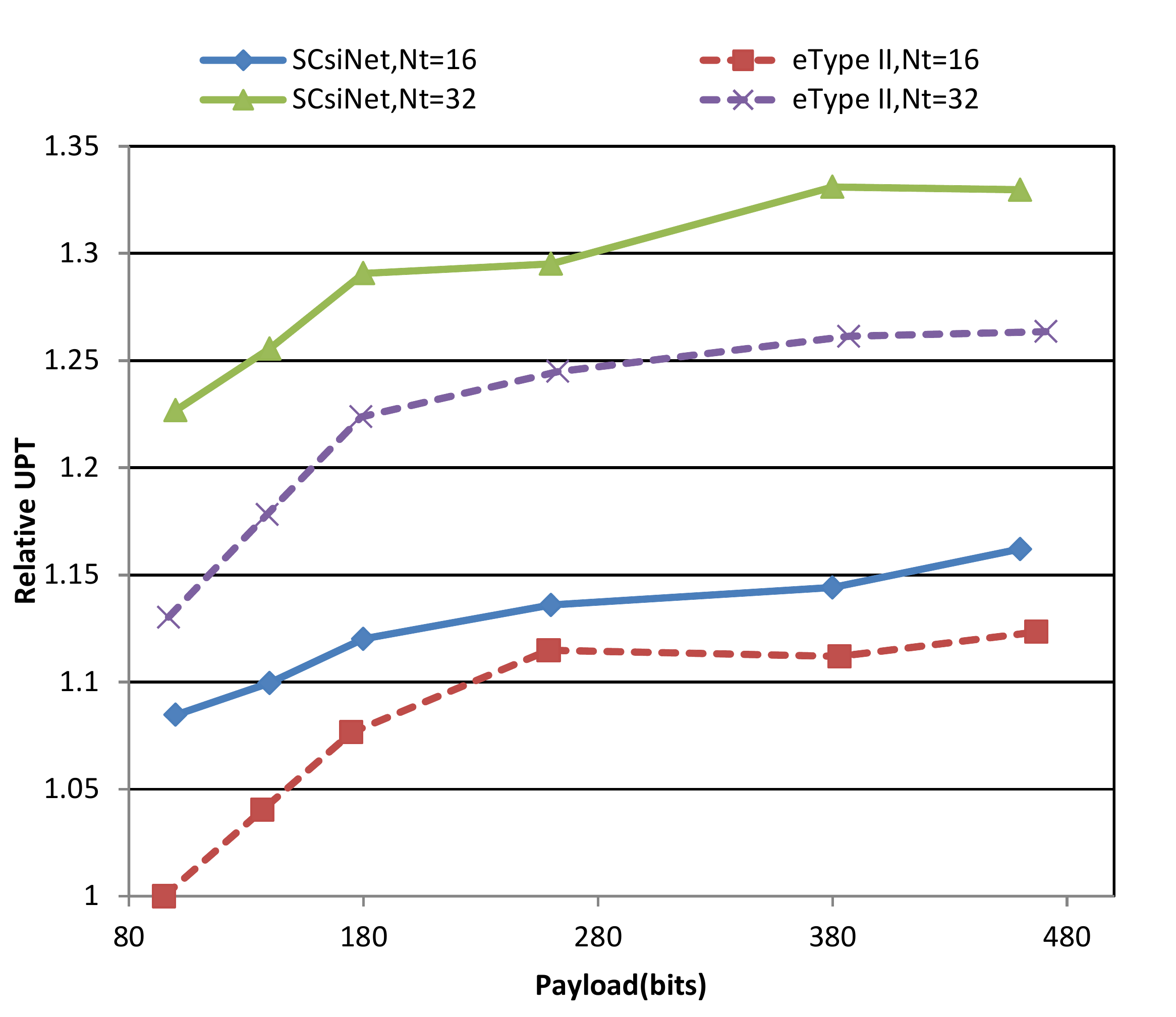}
\caption{UPT performances of SCsiNet and eType II codebook based scheme for adaptive rank and FTP model-1 service.}\label{fig:UPT}
\end{figure}

\begin{table}[htbp]
\centering
\caption{Payload allocations (bits) among 4 layers for different ranks.}
\begin{tabular}{|c|c|c|c|c|}
\hline
\textbf{Configuration}&\textbf{Rank1}&\textbf{Rank2}&\textbf{Rank3}&\textbf{Rank4}\\
\hline
\textbf{1}&40&60,20&40,20,20&40,20,20,20\\
\hline
\textbf{2}&60&80,40&60,40,20&60,20,20,40\\
\hline
\textbf{3}&80&100,60&80,40,40&80,40,20,40\\
\hline
\textbf{4}&120&160,80&100,80,60&120,60,40,40\\
\hline
\textbf{5}&160&220,100&160,120,80&160,100,60,60\\
\hline
\textbf{6}&240&320,140&180,140,120&180,140,80,60\\
\hline
\end{tabular}\label{table:payload_allocation}
\end{table}

In wireless systems, the performance gain of increasing eigenvector feedback accuracy for high ranks
is much lower than that of low ranks, and the probability of scheduling UEs with high ranks is also
much lower with the respect to low ranks. Therefore, allocating more payloads for high ranks than low ranks is not a good choice. In the 5G NR system, feedback payload of rank=2,3,4 is approximately twice compared to rank=1. Thanks to the scalable and flexible framework, the payload allocation among different layers and ranks can be easily realized by SCsiNet. Table.\ref{table:payload_allocation} gives payload allocation among 4 layers for different ranks.

Fig.\ref{fig:UPT} shows UPT performances of SCsiNet and eType II codebook based schemes for $N_t=16,32$ and various payloads. The SCsiNet is deployed into system level simulations with C/C++ interface of deep learning
platform. Rank adaptive schedule and FTP model 1 service are adopted and packet size is 0.5Mbytes.
It can be seen from Fig.\ref{fig:UPT} that the proposed SCsiNet outperforms the eType II codebook based scheme for $N_t=16,32$ antenna number and all payloads. Under the same feedback payload, the proposed SCsiNet can achieve approximately $2\%\sim10\%$ UPT improvement.

\subsection{FLOPs and model parameters}
\begin{table}[htbp]
\centering
\caption{FLOPs ($10^6$) of SCsiNet.}
\begin{tabular}{|c|c|c|c|c|c|c|}
\hline
\textbf{Payload(bits)}&\textbf{20}&\textbf{40}&\textbf{60}&\textbf{80}&\textbf{100}&\textbf{120}\\
\hline
\text{$N_t=16$, encoder}&9.8&9.84&9.87&9.9&9.93&9.96\\
\hline
\text{$N_t=16$, decoder}&9.8&9.84&9.87&9.9&9.93&9.96\\
\hline
\text{$N_t=32$, encoder}&9.9&9.93&9.97&10&10&10.06\\
\hline
\text{$N_t=32$, decoder}&9.9&9.93&9.96&10&10.03&10.06\\
\hline
\textbf{Payload(bits)}&\textbf{140}&\textbf{160}&\textbf{180}&\textbf{200}&\textbf{220}&\textbf{240}\\
\hline
\text{$N_t=16$, encoder}&9.96&10.02&10.1&10.08&10.11&10.14\\
\hline
\text{$N_t=16$, decoder}&9.96&10.02&10.05&10.08&10.11&10.14\\
\hline
\text{$N_t=32$, encoder}&10.06&10.12&10.15&10.18&10.21&10.24\\
\hline
\text{$N_t=32$, decoder}&10.06&10.12&10.15&10.18&10.21&10.24\\
\hline
\textbf{Payload(bits)}&\textbf{260}&\textbf{280}&\textbf{300}&\textbf{320}&&\\
\hline
\text{$N_t=16$, encoder}&10.17&10.21&10.24&10.27&&\\
\hline
\text{$N_t=16$, decoder}&10.17&10.2&10.24&10.27&&\\
\hline
\text{$N_t=32$, encoder}&10.27&10.3&10.34&10.37&&\\
\hline
\text{$N_t=32$, decoder}&10.27&10.3&10.33&10.36&&\\
\hline
\end{tabular}\label{Table:flops}
\end{table}

\begin{table}[htbp]
\centering
\caption{Parameter number ($10^6$) of SCsiNet.}
\begin{tabular}{|c|c|c|}
\hline
\textbf{Model type}&\textbf{Encoder}&\textbf{Decoder}\\
\hline
\text{Total parameters}&2.51&2.52\\
\hline
\end{tabular}\label{table:para_num}
\end{table}

The inference complexity (in terms of floating point operations, FLOPs) of SCsiNet is given in Table.\ref{Table:flops} for all payloads and antenna numbers. And the total parameter number of SCsiNet is given in Table.\ref{table:para_num} for both encoder part and decoder part. Since the structure of SCsiNet is almost symmetrical, both FLOPs and parameter numbers are similar for encoder part and decoder part. It can be seen from \cite{WLiu} that the total parameter number of EVCsiNet is about $4.6\times {10}^{6}$ for only one configuration and rank=1. However, the proposed SCsiNet can support $32$ configurations and multiple ranks with the similar parameter number of $5\times{10}^6$. This significantly reduces the burden of training, transferring and deploying DL models.

\section{Conclusion}
In this paper, a scalable and flexible framework for DL based eigenvector feedback called SCsiNet is proposed to adapt various payloads, ranks and antenna numbers. Simulation results show that the proposed
SCsiNet has similar performances but simple structure compared to configuration-orienting DL schemes, and achieves approximately $2\%\sim10\%$ UPT improvement compared to eType II codebook based schemes.

\section{Appendix}
Simulation assumptions and channel models we adopted in this paper are given in Table.\ref{table:sim_asump}.

\begin{table}[htbp]
\centering
\caption{Simulation assumptions}
\begin{tabular}{|c|c|}
\hline
\textbf{Parameter}&\textbf{Value}\\
\hline
\text{Duplex,Waveform}&\text{FDD,OFDM}\\
\hline
\text{Scenario}&\text{UMa}\\
\hline
\text{Frequency,SCS}&\text{2GHz,15kHz}\\
\hline
\text{Inter-BS distance,}&\text{200m,}\\
\text{BS antenna height,}&\text{25m},\\
\text{BS Tx power}&\text{41dBm}\\
\hline
\text{Channel model}&\text{Accoring to 3GPP TR38.901}\\
\hline
\text{Antenna setup}&\text{32 ports:}$(M,N,P,M_g,N_g,M_p,N_p)$\\
\text{and port layouts}&=(8,8,2,1,1,2,8),\\
\text{at the BS}&\text{(dH,dV)=(0.5,0.8)$\lambda$};\\
&\text{16 ports:}$(M,N,P,M_g,N_g,M_p,N_p)$\\
&=(8,4,2,1,1,2,4),\\
&\text{(dH,dV)=(0.5,0.8)$\lambda$}\\
\hline
\text{Antenna setup}&\text{4Rx:}$(M,N,P,M_g,N_g,M_p,N_p)$\\
\text{and port layouts at UEs}&=(1,2,2,1,1,1,2)\\
&\text{(dH,dV)=(0.5,0.5)$\lambda$}\\
\hline
\text{Bandwidth}&\text{48PRBs}\\
\hline
\text{UE distribution}&\text{80\% indoor (3km/h),}\\
&\text{20\% outdoor (30km/h)}\\
\hline
\text{UE receiver}&\text{MMSE-IRC}\\
\hline
\text{CSI feedback periodicity}&\text{5ms}\\
\hline
\text{Traffic model}&\text{FTP, model 1, 0.5MBytes}\\
\hline
\end{tabular}\label{table:sim_asump}
\end{table}

\end{document}